# Extruded Mg based hybrid composite alloys studied by longitudinal impression creep

## A.K. Mondal, S. Kumar

*Department of Materials Engineering, Indian Institute of Science, Bangalore 560012, India*

*The creep behaviour of a creep-resistant AE42 magnesium alloy reinforced with Saffil short fibres and SiC particulates in various combinations has been examined in the longitudinal direction, i.e., the plane containing random fibre orientation was parallel to the loading direction, in the temperature range of 175– 300 C at the stress levels ranging from 60 to 140 MPa using impression creep test technique. At 175 C, normal creep behaviour, i.e., strain rate decreasing with strain and then reaching a steady state, is observed at all the stresses employed. At 240 C, normal creep behaviour is observed up to 80 MPa and reverse creep behaviour, i.e., strain rate increasing with strain, then reaching a steady state and again decreasing, is observed above that stress. At 300 C, reverse creep behaviour is observed at all the stresses employed. This pattern remains the same for all the composites. The reverse creep behaviour is found to be associated with the fibre breakage. The stress exponent is found to be very high for all the composites. However, after taking the threshold stress into account, the stress exponent varies from 3.9 to 7.0, which suggests viscous glide and dislocation climb being the dominant creep mechanisms. The apparent activation energy Qc was not calculated due to insufficient data at any stress level either for normal or reverse creep behaviour. The creep resistance of the hybrid composites is found to be comparable to that of the composite reinforced with 20% Saffil short fibres at all the temperatures and stress levels investigated.*

## 1. Introduction

   In recent times, there has been a great upsurge in using magnesium for structural applications, the demand mainly coming from automotive industry owing to environmental concerns, increasing safety and comfort levels, a significant improvement in the corrosion resistance of high purity magnesium alloys, rising fuel prices and lowering of prices of primary magnesium metal. For powertrain components, several creep-resistant magnesium alloys have been developed [1]. However, for temperatures above 200 C, metal matrix composites (MMCs) need to be developed. Particulate reinforced Mg-MMCs might actually deteriorate the creep properties [2,3]. Therefore, short fibre reinforced MMCs have to be developed for these applications. However, they are expensive and have anisotropic properties. Therefore, hybrid composites (HC) would be the best choice. The partial replacement of expensive short fibres by cheap particulates reduces the cost as well as anisotropy. The addition of particulates also helps in keeping the fibres apart, which improves mechanical properties [4,5]. Friend et al. observed that the partial replacement of Saffil short fibres by SiC particulates in 7039 Al alloy-based hybrid MMCs did not affect the strength and elastic moduli but improved fracture toughness [6]. Park observed an improvement in wear resistance of 6061 Al alloy-based hybrid MMC reinforced with Al2O3 short fibres and SiC whiskers than that in MMCs reinforced either with Al2O3 short fibres or SiC whiskers alone [7]. However, there have been only a few studies on Mg alloy-based hybrid MMCs [8–16].

   Creep studies are very important for MMCs being developed for engine components, as they are subjected to high temperature in these applications. These properties of MMCs are currently determined in conventional tensile and compressive creep tests [15–18]. There has been only one study on the creep behaviour of Al alloy-based MMCs using indentation creep [19]. However, to the best of our knowledge, there has been no report on the creep behaviour of MMCs using impression creep, though it has been identified as a versatile method for determining creep properties of materials long ago [20,21]. In the present study, creep behaviour of a creep-resistant AE42 alloy reinforced with 20 vol.% Saffil short fibres as well as various volume fractions of Saffil short fibres and SiC particulates has been studied in the longitudinal direction, i.e., the plane containing random fibre orientation was parallel to the loading direction, using impression creep technique.

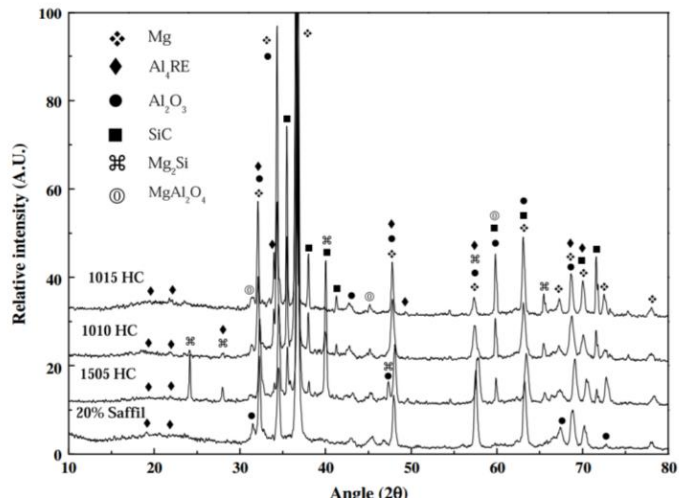

**Fig. 1.** XRD patterns obtained from all the composites.

## 2. Experimental procedure

The magnesium alloy AE42 (Mg–4.0 wt.% Al–2.0 wt.% Rare Earth (RE)–0.2 wt.% Mn) based composites are used in the present investigation. The alloy is reinforced with Saffil short fibres having 3–8 lm diameter and 200 lm length and SiC particulates having a maximum diameter of 40 lm, in various combinations. The various combinations are 20 vol.% Saffil short fibres, 15 vol.% Saffil short fibres and 5 vol.% SiC particulates, 10 vol.% Saffil short fibres and 10 vol.% SiC particulates, and 10 vol.% Saffil short fibres and 15 vol.% SiC particulates. For convenience, the composites are henceforth designated as 20% Saffil, 1505 HC, 1010 HC and 1015 HC, respectively. All the materials were fabricated by squeeze casting process and the detailed procedure has been discussed elsewhere [12]. The solidified composite contains a dense microstructure with Saffil fibres in a planar random array. The creep samples having a dimension of 10 mm * 10 mm * 7 mm were machined from the block by spark erosion technique. All the materials were tested in the longitudinal direction, i.e., the plane containing random fibre orientation was parallel to the loading direction. The creep tests were carried out using impression creep technique, in which a cylindrical indenter is impressed on the sample and the depth of penetration is monitored as a function of time. The depth of penetration is then converted to strain by dividing it with the diameter of the indenter used. The detailed procedure about the technique can be found elsewhere [21]. In the present investigation, a tungsten carbide indenter of 1.5 mm diameter was used and the tests were conducted in air in the temperature range of 175–300 C at stress levels ranging from 60 to 140 MPa. The test temperature was maintained within an accuracy of ±2 C during the creep tests. Microstructures of all the squeeze cast samples of the AE42 alloy-based composites were examined before and after creep tests using a FEI scanning electron microscope (SEM) equipped with an energy dispersive X-ray spectroscopy (EDS) analysis. The samples for microstructural analysis were prepared by standard metallographic technique. For etching, a solution of 100 ml ethanol, 10 ml acetic acid, 6 ml picric acid and 20 ml of distilled water was used. The phases present were identified by X-ray diffraction (XRD) (model JEOL, JDX 8030) using Cu Ka (k = 1.541 Å) radiation.

## 3. Results and discussions

### 3.1. As-cast microstructure

Fig. 1 shows the XRD patterns of all the squeeze cast AE42 alloybased composites. It is evident that all the as-cast composites consist of a-Mg peaks along with the peaks corresponding to Al4RE phase. The phases Al11RE3 and Al4RE are considered to be same and are conventionally denoted as Al4RE phase in the literature [22]. There was no peak corresponding to the Mg17Al12 phase. Thus, the low melting point phase Mg17Al12, which is present in Mg–Al alloys, is completely suppressed with the addition of RE in the present alloy, which is beneficial for creep resistance. Powell et al. [22] have also observed only a-Mg and Al4RE phase in a die-cast AE42 alloy. Moreno et al. [23] examined the microstructure of Mg–3.7 wt.% Al–2.69 wt.% RE–0.21 wt.% Mn alloys using transmission electron microscope and reported that both a-Mg and Al4RE phase exist in the alloy. Wu et al. [24] also observed Al11RE3 phase in the Mg–3.9 wt.% Al–2.0 wt.% RE alloy. Besides aMg and Al4RE phase, in case of 20% Saffil composite, Al2O3, and in case of hybrid composites Al2O3 and SiC are the main phases present in the as-cast condition. Apart from these phases, some intermetallic phases, like Mg2Si and MgAl2O4 are also present in all the composites investigated, which are produced by the reactions of Mg with SiC and Al2O3, respectively. Fig. 2(a–d) shows the SEM micrographs of all the composites in the plane normal to the indenter. In this plane, essentially the cross section of the Saffil fibres, with only few fibres in the longitudinal direction, are observed. The solidified composites contain a dense microstructure with Saffil fibres and SiC particulates being uniformly distributed. The interfacial integrity between reinforcements and matrix, as observed at high magnification (figure not shown), was good, which indicates good wetting of reinforcements by Mg-matrix during processing. The Al4RE phase having bright contrast was observed attached to the fibre as well as particulates (shown by A in Fig. 2(a)). The reaction product Mg2Si was observed attached to the SiC particulates (shown by B in Fig. 2(d)) in hybrid composites.

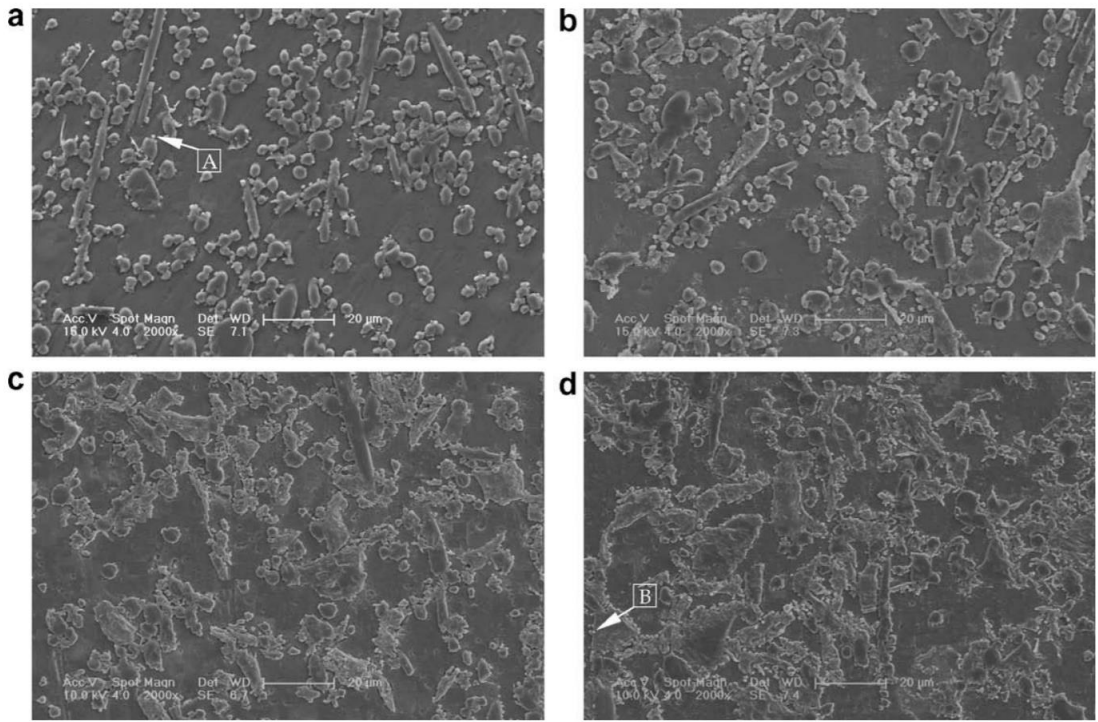

Fig. 2. SEM micrographs in the plane normal to the indenter of the composites (a) 20% Saffil (b) 1505 HC (c) 1010 HC and (d) 1015 HC.

## 3.2. Nature of the creep curves

Fig. 3(a) shows the variation of time dependent strain with time in impression creep test for all the composites tested at 175 C at different stress levels in the longitudinal direction. All curves exhibit a distinct primary creep regime followed by a steady state creep, as has been observed in several other Mg-MMCs [15–18]. From the slope of the second part of the curve a nearly constant steady state creep rate can be determined. The strain rate vs. strain curves derived from the strain vs. time curves shown in Fig. 3(a) is plotted in Fig. 3(b). It is evident from the figure that at the first stage, the strain rate is gradually decreased and in the second stage, a steady state strain rate is achieved. The creep curves of all the composites tested at temperature of 175 C at all stresses used and at temperature of 240 C up to 80 MPa stress, exhibit behaviour similar to that observed in conventional creep. Henceforth, we shall refer this creep behaviour as normal creep behaviour. Creep curves at higher stresses and temperatures differ signifi- cantly from those obtained at lower stresses and temperatures shown in Fig. 3 for all the composites. Fig. 4(a) shows the variation of time dependent strain with time in impression test creep, obtained at 240 C, at stresses above 80 MPa and Fig. 4(b) shows the strain rate vs. strain curves derived from Fig. 4(a) for two hybrid composites as examples. Three stages can clearly be distinguished on each of the curves. The first stage (I) is characterized by a gradually increasing strain rate, the second stage (II) by a constant strain rate, and in the third stage (III) the strain rate is gradually decreased. A similar behaviour was observed at 300 C for all the stresses used. The impression creep behaviour of composites at higher stresses and temperatures is just reverse to the normal creep behaviour discussed above. Henceforth, we shall refer this creep behaviour as reverse creep behaviour. There has been only one study in indentation creep of Al alloy reinforced with 13.9 vol.% Saffil short fibre by Cseh et al. [19]. These authors carried out the creep tests only at higher stresses and temperatures and observed a reverse creep behaviour, as observed in the present investigation, at all the stresses and temperatures employed. A summary of the creep results, with respect to normal (N) and reverse (R) behaviour, observed in the present investigation, is shown in Table 1. At 175 C, normal behaviour is observed at all the stresses employed. At 240 C, normal behaviour is observed up to 80 MPa and reverse behaviour above that stress. At 300 C, reverse behaviour is observed at all the stresses employed. This pattern remains the same for all the composites. In order to understand the reverse creep behaviour, a detailed microstructural analysis of the composites at each stage of the creep test was carried out. The representative microstructures of the 20% Saffil composite, tested at 300 C temperature and 100 MPa stress, using a 1.5-mm diameter indenter corresponding to stages I and III are shown in Fig. 5(a–b), respectively. These microstructures are taken at a plane below the indenter after cutting the specimen in two halves along the indenter direction. It should be noted that these planes are perpendicular to the planes shown in Fig. 2 and exhibit essentially the longitudinal section of 3252 A. the fibres oriented randomly along with few transverse section of fibres. It can be seen in Fig. 5(a) that the failure of the fibres (marked by arrow) appears first in the vicinity of the rim of the indenter, where, according to theoretical considerations, the shear stress concentration is highest [25]. Cseh et al. also concluded a similar behaviour with the help of ANSYS finite element analysis program [19]. They reported that the plastic deformation zone extends gradually from the edge to underneath the indenter in their model, which is in good agreement with the observed microstructure in the present investigation.

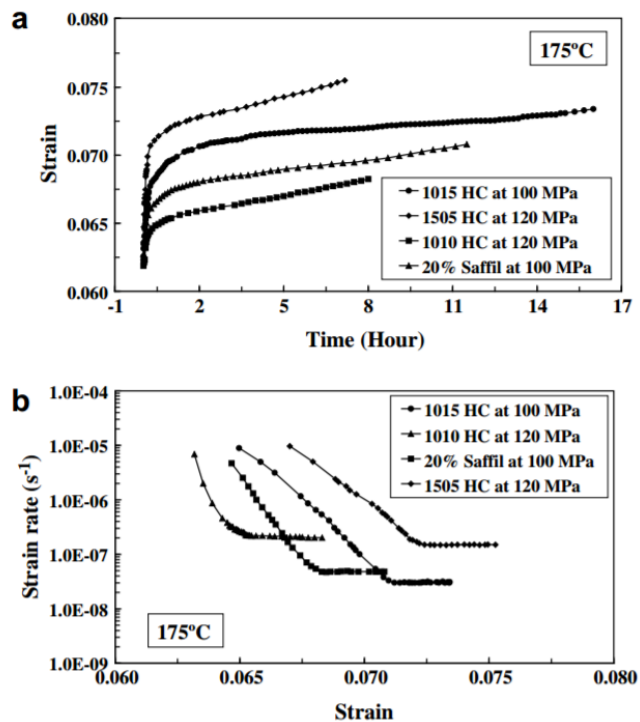

Fig. 3. (a) Variation of strain with time for all the composites tested at 175 C and various stress levels in the longitudinal direction and (b) the strain rate vs. strain curves corresponding to (a).

The decrease in strain rate during stage I in normal creep behaviour is due to strain hardening of the matrix, as observed in Fig. 3(b). An increase in strain rate during stage I in reverse creep behaviour observed in Fig. 4(b) can be explained as follows: initially the matrix deforms slowly with the generation and movement of dislocations. The movement of dislocations is hindered by the rigid network of the fibres and dislocation pile-ups are formed. As the formation of pile-ups is continued, the local stress around the fibres is increased accordingly. The increase in local stress concentration leads to the fibre fracture. This helps the dislocations start moving, inducing failure of further fibres, thereby accelerating the deformation. This has also been confirmed by plotting the first part of the impression creep curve at higher resolution, as shown in Fig. 6. The careful observation of Fig. 6 shows steps of several microns height, each of which corresponds to a fi- bre fracture in the fibre network. Similar type of steps in the deformation curve, around the minimum creep rate, during tensile creep tests on fibre reinforced MMCs has also been observed and attributed to fibre fracture in a previous study [26]. In stage II, a steady state creep rate is attained. It suggests that fibre fracture has completed and microstructure does not change in this region with time. The deformation zone moves in front of the indenter and a constant strain rate is attained, as it is the characteristic of steady state secondary creep. Similar observations about the propagation of the deformation zone have been reported previously for unreinforced age-hardenable Al alloys using impression creep [27]. Fig. 5(b) shows the microstructure of the composite in the third stage (III), where the creep rate is gradually decreased. Immediately beneath the indenter is the elastic region, which is the characteristic of the indentation [25]. At this stage, the broken fibres in the plastic deformation zone are clogged, i.e., all broken fibres are totally entrapped and segregated below and surrounding the indenter so that the further movement of the indenter is slowed down. The broken fibres also change their orientation and tend to align in the direction of flow within the bulk material, as represented by a kind of flow pattern shown between the two dotted curves in Fig 5(b). This effect can be ascribed to the extensive shear deformation taking place below the indenter, as shown by the results of the FEM analysis by Cseh et al. [19]. This is in contrast to the behaviour in conventional tensile creep tests where multiplefracture of fibres leads to an accelerating tertiary creep leading to the rupture of the sample [26].

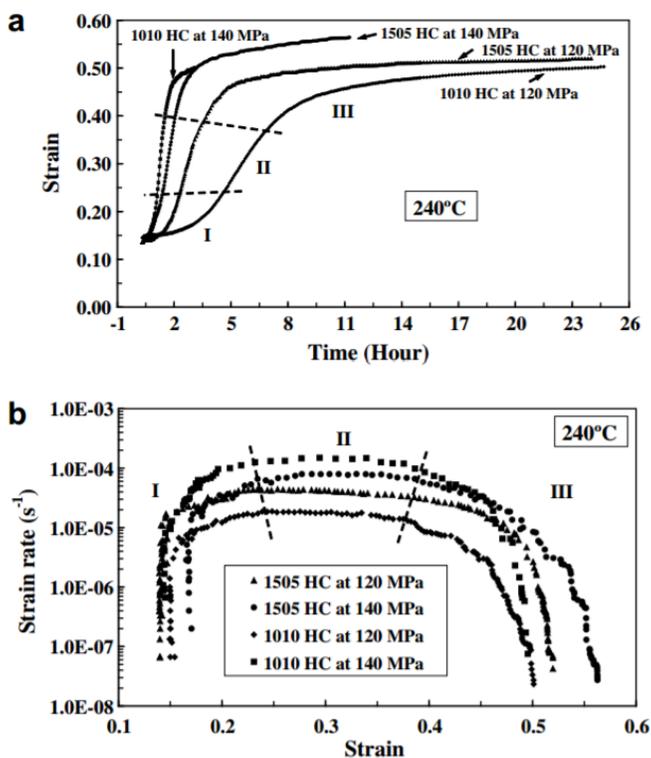

Fig. 4. (a) Variation of strain with time for two hybrid composites tested at 240 C above 80 MPa in the longitudinal direction and (b) the strain rate vs. strain curves corresponding to (a).

### 3.3. Evaluation of steady state creep mechanism

The creep rate was calculated from the steady state region of the strain rate vs. strain curves for both the normal and reverse creep behaviour i.e., from the second stage in each case. The stress dependence of the steady state creep rate at different temperatures in double logarithmic scale for all the composites has been plotted and the representative plot for 1010 HC is shown in Fig. 7. The values of the apparent stress exponent, na, were calculated by fitting straight line to the data, at a fixed temperature, for all the composites and are summarized in Table 2. As there is a transition from normal to reverse creep behaviour at 240 C above 80 MPa stress, two separate straight lines have been fitted in two regions and the values of the apparent stress exponent, na, has also been calculated separately.

It is well-documented that for metals and solid solution alloys, the dependence of steady state creep rate, $\dot{\varepsilon}_s$, on the applied stress, $\sigma$, at intermediate stresses, can be represented as a power law of the applied stress, as follows [28,29]

$$\dot{\varepsilon}_s = \frac{ADGb}{kT}\left(\frac{\sigma}{G}\right)^n$$

where A is a dimensionless constant, D is the appropriate diffusion coefficient [¼ D0exp Qc RT Þ; where D0 is a frequency factor, Qc is the activation energy for creep, R is the universal gas constant and T is the absolute temperature], G is the shear modulus, b is the Burgers vector, k is Boltzmann's constant and n is the stress exponent. For pure metals (such as Al) and solid solution alloys of class II (metal class) (such as Al–5 wt.% Zn [30]), n 5 whereas, for solid solution alloys of class I (alloy class) (such as Al–Cu alloys [31]), n can be 3 or 5 depending on the experimental conditions and material parameters [32,33]. As can be seen from Table 2, the values of the apparent stress exponent 'na, obtained in the present investigation are quiet high and are in the range from 7.2 to 16.0. Such high values of the apparent stress exponent, na, cannot be explained based on the existing deformation models. The discontinuously reinforced metal matrix composites (DRMMCs) are known to exhibit high values of apparent stress exponent, na [17,34–36], which has been attributed to the existence of a threshold stress, r0. The threshold stress is de- fined as a lower limiting stress below which no measurable strain rate can be achieved [37].

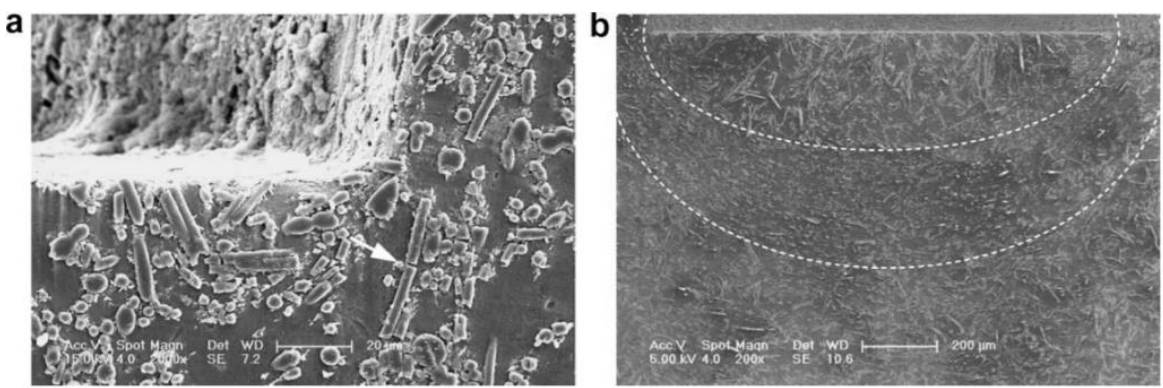

Fig. 5. SEM micrographs of the 20% Saffil composite tested at 300 C and 100 MPa corresponding to (a) stage I and (b) stage III.

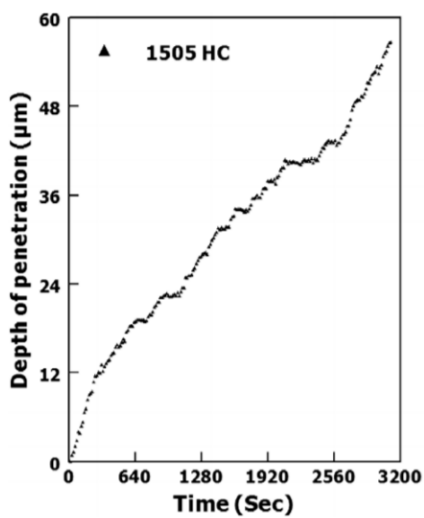

Fig. 6. The first part of the impression creep curve at higher resolution, obtained for 1505 HC at 240 C and 140 MPa.

The reason behind the origin of the threshold stress is not yet fully clear. Three deformation models have been proposed to explain the origin and give the magnitude of threshold stress. According to these models, the threshold stress is equal to (a) the stress required to cause dislocation bowing between particles (Orowan stress) [38], (b) the stress required to detach a dislocation from the particles after climb is over [39,40] and (c) the additional back stress necessary to create an additional dislocation line length as the dislocation climbs over a particle (local climb). When a threshold stress is present, the effective stress, $\sigma_e$ ($=\sigma-\sigma_0$) replaces the applied stress, $\sigma$, and the steady state creep rate, $\dot{\varepsilon}_s$, is expressed through a relationship of the form $\dot{\varepsilon}_s \frac{1}{4} \frac{ADGb}{kT} \left(\frac{\sigma-\sigma_0}{G}\right)^n \eth2\bkP$ where $\sigma_0$ is the threshold stress and the other parameters have been defined earlier. The estimation of the magnitude of threshold stress, $\sigma_0$, is done in several ways. The experimental data at a single temperature are generally plotted as $\dot{\varepsilon}_s^{1/n}$ against $\sigma$ in double linear scales with known values of n [41] (n = 3 for creep controlled by viscous glide processes [32], n = 5 for creep controlled by high temperature climb of dislocations (lattice diffusion) [29,42], n = 7 for creep controlled by low temperature climb (core diffusion) [43] and n = 8 for lattice diffusion controlled creep with a constant structure [44]). If the creep rate of the composite obeys Eq. (2) and if $\sigma_0$ is constant for each test temperature (independent of the applied stress), in case of the correct choice of n, the data points of the above plot will fit into a straight line whose extrapolation to zero strain rate gives the value of threshold stress, $\sigma_0$. Li and Langdon developed a more simple method for estimating the threshold stress in DRMMCs [37]. They criticised the priori selection of the values of stress exponent, n, as described above. They consider a creep rate of $10^{10}$ s1 as it is the lowest measurable creep rate in lab and this corresponds to a creep deformation of approximately been attributed to the attractive interactions between mobile dislocations and a dispersion of fine incoherent oxide particles introduced during the atomization process [34,45–48]. However, the composites in the present investigation were prepared by squeeze casting and do not contain these oxide particles. Li and Langdon [49] have attributed the temperature dependent threshold stress in the AZ91 Mg alloy reinforced with alumina fibres prepared by squeeze casting technique to the presence of $Mg_{17}Al_{12}$ precipitates in the matrix. The AE42 Mg alloy used in the present investigation also contains $Al_4RE$ precipitates. However, whether these precipitates can give rise to a temperature dependent threshold stress needs to be further investigated. After incorporating the calculated threshold stresses, $\sigma_0$, Fig. 9 shows the representative plot for 1010 HC in double logarithmic plots of the steady state creep rate against the effective stress, $\sigma_e$. From the figure, the values of the true stress exponent, $n_0$, have been calculated and are presented in Table 2. The values of the true stress exponent, $n_0$, obtained in the present investigation match well with that reported on the same materials under conventional compression creep tests by Arunachaleswaran et al. [15] and Dieringa et al. [17]. The values of stress exponent, $n_0$, varies from 3.9 to 7.0, which suggests the viscous glide as well as climb of dislocation being the dominant creep mechanism [29,32,42]. Viscous glide or a combination of viscous glide with dislocation climb were reported to be the dominant creep mechanisms in Mg alloy-based composites [15,17,18]. The apparent activation energy $Q_c$ cannot be calculated from the above data, since we do not have more than two data at any stress level either for normal or reverse creep behaviour. The normal and reverse creep data should not be mixed together for the calculation of $Q_c$, since they represent different microstructural states.

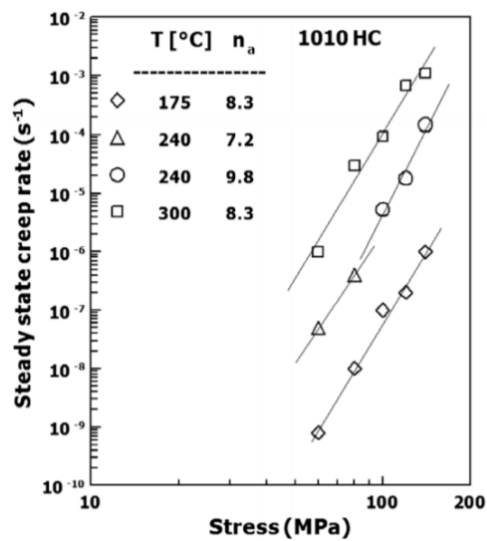

Fig. 7. The stress dependence of the steady state creep rate for 1010 HC.

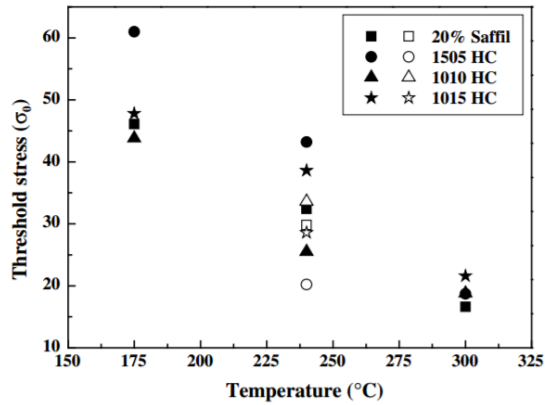

Fig. 8. The calculated threshold stress, r0, as a function of temperature for all the composites.

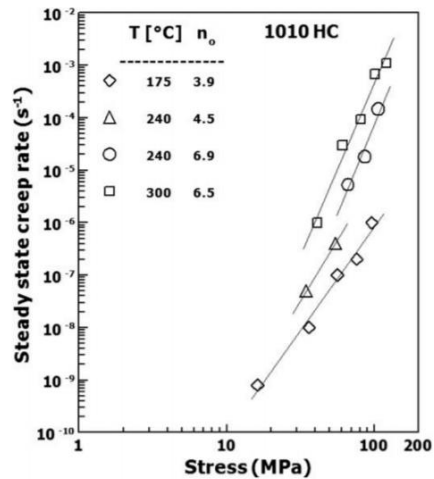

Fig. 9. The steady state creep rate against the effective stress for 1010 HC.

### 3.4. Effect of reinforcement volume fraction on creep rate

The steady state creep rate at a fixed temperature for all the composites has been compared and is shown in Fig. 10(a–c). It is evident that the creep rate of the hybrid composites are comparable with the composite reinforced only with 20% Saffil short fibres. So, from the commercial point of view, the use of the hybrid composites, replacing small volume fraction of expensive Saffil short fi- bres by cheap SiC particulates, is beneficial. The short fibre reinforced Mg-MMCs generally increase the creep resistance [18,50] by two different mechanisms [51-58]. First way is the load transfer effect in which part of the external load is carried by the reinforcement. Second way is a substructural strengthening effect due to features such as the increased dislocation density, which is present in the composite due to the thermal mismatch between the matrix and the reinforcement. However, the particulate reinforced Mg-MMCs might actually deteriorate the creep properties [2,3]. Mordike et al. [2] suggested that the slip process at grain and particulate/matrix interfaces as well as diffusion caused a sudden increase in secondary creep rate, which lead to fracture of the specimens after a short period of loading in a 15 vol.% SiC particulates reinforced QE22 alloy. In another investigation, Moll et al. [3] attributed the poorer creep resistance of the QE22 alloy reinforced with 15 vol.% SiC particulates, as compared to the monolithic QE22 alloy, to the additional contribution of boundary and interfacial sliding. The comparable creep resistance of the hybrid composites with that of the 20% Saffil composite in the present investigation may be attributed to the reaction products obtained from the interfacial reaction between Mg matrix and SiC particulates. The interfacial reaction between Mg matrix and SiC particulates yields fine Mg2Si particles on the reinforcements/matrix interface, as shown in Figs. 1 and 2. The presence of Mg2Si may act as an additional reinforcement in case of hybrid composites and increase the creep resistance. The addition of Mg2Si particulates as reinforcement in the Mg alloy has been reported to improve mechanical properties of Mg-based composite [59-68]. Arunachaleswaran et al. [15] also found beneficial effect of the phase Mg2Si on the creep resistance of their composites. The larger thermal mismatch between magnesium and SiC could also give rise to a large dislocation density and hence an increase in creep resistance of the hybrid composites. Thus, the creep resistance of the hybrid composites remains comparable to the composite reinforced with Saffil fibre alone.

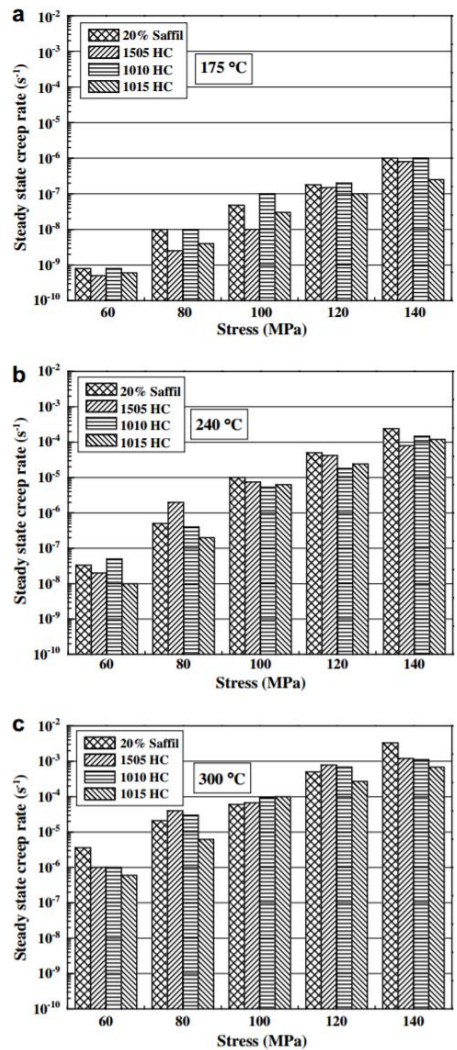

Fig. 10. The comparison of the creep rate for all the composites at temperature of (a) 175 C, (b) 240 C and (c) 300 C.

## 4. Conclusions

In the present investigation, the creep behaviour of a creepresistant AE42 magnesium alloy reinforced with Saffil short fibres and SiC particulates in various combinations has been examined in the longitudinal direction, in the temperature range of 175–300 C and stress levels ranging from 60 to 140 MPa using impression creep test technique. The conclusions arising out from the present study are as follows:

1. At 175 C, normal creep behaviour is observed at all the stresses employed. At 240 C, normal creep behaviour is observed up to 80 MPa and reverse creep behaviour above that stress. At 300 C, reverse creep behaviour is observed at all the stresses employed. This pattern remains the same for all the composites.

2. The reverse creep behaviour is observed due to extensive fibre breakage.

3. Viscous glide as well as climb of dislocation is found to be the dominant creep mechanism in the stress and temperature range employed for all the composites. 4. The creep resistance of the hybrid composites is found to be comparable to that of the composite reinforced only with 20 vol.% Saffil fibres at all the temperatures and stresses tested. So, from the commercial point of view, the use of the hybrid composites, replacing a part of the expensive Saffil short fibres by cheap SiC particulates, is beneficial.